\documentclass[aps,prb,print,groupedaddress,twocolumn]{revtex4-1}
\usepackage{graphicx}
\usepackage{dcolumn}
\usepackage{bm}
\usepackage{subfigure}
\usepackage{color}
\begin{document}
%\setpagewiselinenumbers
%\modulolinenumbers[1]
%\linenumbers

\title{Anomalous Hall Effect and Spin Fluctuations in Ionic Liquid Gated \\ SrCoO$_3$ Thin Films}%
\author{Ding Zhang$^{1,2}$}
\email{dingzhang@mail.tsinghua.edu.cn}
\author{Hiroaki Ishizuka$^{3}$}
\author{Nianpeng Lu$^{1,2}$, Yujia Wang$^1$, Naoto Nagaosa$^{3,4}$, Pu Yu$^{1,2,4}$}
\email{yupu@mail.tsinghua.edu.cn}
\author{Qi-Kun Xue$^{1,2}$}
\affiliation{$^1$State Key Laboratory of Low Dimensional Quantum Physics and Department of Physics, Tsinghua University, Beijing 100084, China\\
$^2$Collaborative Innovation Center of Quantum Matter, Beijing 100084, China\\
$^3$Department of Applied Physics, University of Tokyo, Bunkyo, Tokyo 113-8656, Japan\\
$^4$RIKEN Center for Emergent Matter Science (CEMS), Wako, Saitama 351-0198, Japan\\
}

\date{\today}%

\begin{abstract}
The recent realization of epitaxial SrCoO$_3$ thin films has triggered a renewed interest in their electronic, magnetic, and ionic properties. Here we uncover several unusual magneto-transport properties of this compound, suggesting that it hosts persistent spin fluctuation down to low temperatures. We achieve the metallic SrCoO$_3$ with record-low resistivity from insulating SrCoO$_{2.5}$ by the ionic liquid gating. We find a linear relationship between the anomalous Hall resistivity and the longitudinal resistivity, which cannot be accounted for by the conventional mechanisms. We theoretically propose that the impurity induced chiral spin fluctuation gives rise to such a dependence. The existence of spin fluctuation manifests itself as negatively enhanced magneto-resistance of SrCoO$_3$ when the temperature approaches zero. Our study brings further insight into the unique spin state of SrCoO$_3$ and unveils a novel skew scattering mechanism for the anomalous Hall effect.
\end{abstract}

\maketitle
\section{Introduction}
Transition metal cobaltites are a family of compounds in which the Hund's rule and the crystal field splitting compete fiercely~\cite{Khomskii14}. The process of maximizing the total electronic spin, which is favorable for lowering the exchange energy, gets heavily penalized because of loading electrons onto the $e_g$ orbitals.  The outcome of this competition may be neither a high spin state--when the Hund's rule dominates, nor a low spin state--if the crystal field splitting is large.  Instead, an intermediate spin state can emerge, with its exemplary manifestation in a cubic perovskite--SrCoO$_3$ ~\cite{Potze95,Zhuang98,Kunes12,Hoffmann15}. Recently, single crystals and epitaxial thin films of SrCoO$_3$ become available~\cite{Long11,HNLee13a,HNLee13b}. In contrast to polycrystalline samples studied earlier~\cite{Balamurugan06}, the epitaxial growth of thin films not only stabilizes the perovskite phase but also allows for substrate engineering~\cite{JHLee11}. They are of great importance for room-temperature multiferroic devices, given the Curie temperature of SrCoO$_3$ being at around 300~K and the Neel temperature of SrCoO$_{2.5}$ exceeding 500~K. The epitaxial thin films also possess a more efficient topotactic transformation from SrCoO$_{2.5}$ to perovskite SrCoO$_3$. Conventionally, such a conversion is achieved either by electrolyte induced long time oxidation~\cite{Bezdicka93} or through annealing at high temperatures and high oxygen pressures~\cite{Toquin06}. In thin films, however, this conversion occurs at much less demanding conditions, i.e. shorter time periods, lower temperatures, and reduced oxygen pressures~\cite{HNLee13a,HNLee13b}. Lately, this transformation has been demonstrated by an electric-field controlled process at room temperature~\cite{Ichikawa12,Tambunan14,NPLu17}.

The structural transitions and magnetic ordering in strontium cobaltites have been studied extensively by employing, for example, the X-ray spectroscopy or magnetic susceptibility measurement. The transport properties of SrCoO$_3$ thin films, however, remain largely unexplored. Such an investigation may shed light on the strongly correlated nature~\cite{Balamurugan06} and unusual magnetic anisotropy of this compound~\cite{Long11}. For example, a possible spin glass state was identified in La$_{1-x}$Sr$_x$CoO$_3$ upon investigating its unusual anomalous Hall resistivity~\cite{Onose06}. SrFeO$_3$, a close cousin of SrCoO$_3$, displays multiple helimagnetic phases at low temperatures. These exotic phases manifest themselves in the magneto-resistivity as kinks and hysteretic jumps~\cite{Ishiwata11,Chakraverty13}.  It is therefore of interest to investigate the transport properties of SrCoO$_3$ thin films, given its unique spin state.

Here in this paper, we carry out a systematic magneto-transport study on SrCoO$_3$ thin films down to low temperatures and reveal the existence of persistent spin fluctuation. Through ionic liquid gating~(ILG), we obtain the metallic SrCoO$_3$ with record-low resistivity values from the insulating SrCoO$_{2.5}$. Surprisingly, the anomalous Hall resistivity ($\rho_{AH}$) of SrCoO$_3$ grows linearly as a function of $(\rho_{xx}-\rho_0)$, where $\rho_{xx}$ is the longitudinal resistivity and $\rho_0$ the residual resistivity. We propose theoretically that this behavior is a consequence of a novel type of skew scattering that stems from spin fluctuation with impurity-induced local inversion-symmetry breaking. The scenario of spin fluctuation is supported by the experimentally observed negative magneto-resistance~(MR) in SrCoO$_3$. The MR exhibits a parabolic shape at low magnetic fields and a linear behavior at high fields. Intriguingly, it gets enhanced with a decreasing temperature, well below the Curie transition temperature. After ruling out mechanisms including the surface scattering, anisotropic effect, domain-wall effect and weak localization, we show that the high field negative MR can be reproduced theoretically by considering spin fluctuation. Our work demonstrates that SrCoO$_3$ not only is of importance for applications but also hosts quantum properties that could enrich our understanding on the anomalous Hall effect (AHE).

\section{Sample preparation}

Thin films of SrCoO$_{2.5}$ were grown on (LaAlO$_3$)$_{0.3}$-(SrAl$_{0.5}$Ta$_{0.5}$O$_3$)$_{0.7}$ (001) substrate by a home-designed pulsed laser deposition system. The growth temperature is 750~$^\circ$C with the oxygen pressure of 100~mTorr. The laser energy (KrF, $\lambda$=248~nm) was set at 1.2~J/cm$^2$ with a frequency of 2~Hz. After the growth, samples were cooled down to room temperature with a rate of 5~$^\circ$C/minute. The sample quality was confirmed by X-ray diffraction as well as atomic force microscopy.

Our device under investigation is schematically shown in Figs.~1~(a) and 1~(b). Gold pads were evaporated on the samples as contacts. We carved out the Hall bar structure mechanically. Samples were then immersed together with a Pt counter-electrode into the ionic liquid (DEME-TFSI)~\cite{NPLu17,Ueno14,Yuan10,Jeong13}. The electrochemical reaction and subsequent magneto-transport investigations were carried out in a physical property measurement system (Quantum Design PPMS-9T) with standard lock-in techniques (typically $I_{AC}=1$~$\mu$A, 13~Hz). Pure oxygen gas was filled into the sample chamber to ensure proper oxidization and was later pumped out at around 150~K to avoid the hazardous icing.

As demonstrated in our previous study~\cite{NPLu17}, we can tune from the SrCoO$_{2.5}$ to SrCoO$_3$ through the ILG induced oxygen ion injection. The pristine SrCoO$_{2.5}$~\cite{Munoz08} contains oxygen vacancy chains that run along the [1-10] direction [hexagonal hollow sites in Fig.~1~(b)]. By applying a negative voltage (about -2.5~V) to the gate, oxygen ions can be driven into the sponge-like SrCoO$_{2.5}$ and fill the vacancies to form high-quality SrCoO$_3$~\cite{NPLu17}. The reaction rate is controlled by gating temperature and duration. We achieve fully metallic samples with record-low resistivity values [Fig. ~1~(c)], compared with the previously reported values of the single crystalline bulk~\cite{Long11} and thin films~\cite{HNLee13a}. It indicates high crystalline quality and very low oxygen deficiency: $x\approx3$ in SrCoO$_x$~\cite{HNLee13a,Balamurugan06}.

\begin{figure}
\centering
\includegraphics[width=86mm]{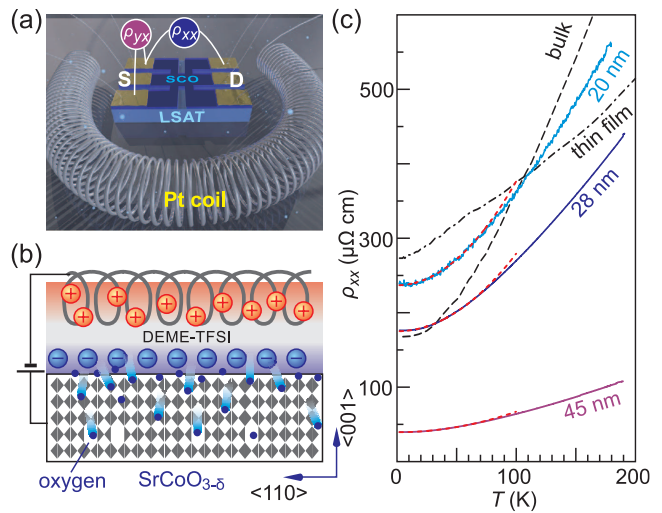}
\caption{(a)~Schematic drawing of the ILG device with the sample in a Hall bar geometry. The Pt coil is the counter-electrode. (b)~Sketch of the strontium cobaltite thin film in contact with the ionic liquid (DEME-TFSI).  (c)~Resistivity as a function of temperature for three gated samples with different thicknesses. Dotted curves are parabolic fittings. The dash (dash-dot) curves represent resistivity of bulk single crystal (thin film) SrCoO$_{3-\delta}$ compounds reported previously~\cite{Long11,HNLee13a}.}
\end{figure}

\section{Anomalous Hall effect}
\subsection{Experiment}

\begin{figure}
\centering
\includegraphics[width=86mm]{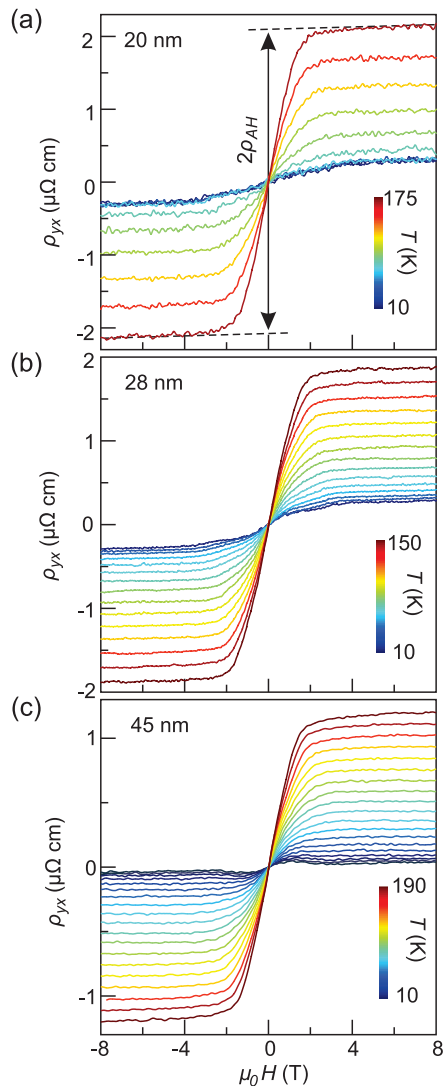}
\caption{Hall resistivity of three samples with different thicknesses at a set of temperatures [(a): $T=$10, 15, 20, 25, 50, 75, 100, 125, 150, 175 K; (b): 10 to 150 K in a step of 10 K; (c): 10 to 190 K in a step of 10 K]. Dashed lines in (a) illustrate the Hall slopes. Each curve is obtained by carefully removing the contribution from the longitudinal resistivity: $\rho_{yx}(\mu_0 H)=\left[\rho_{\rightarrow}(\mu_0 H)-\rho_{\leftarrow}(-\mu_0 H)\right]/2$, where  $\rho_{\rightarrow}(\mu_0 H)$ and $\rho_{\leftarrow}(-\mu_0 H)$ are two Hall traces obtained by sweeping from negative to positive fields and from positive to negative fields, respectively. Linear fits to $\rho_{yx}(\mu_0 H)$ at high fields ($|\mu_0 H|>5$~T) are extrapolated to zero field and the average between the absolute values of the two intercepts is defined as $\rho_{AH}$ [as indicated by the arrows in panel (a)].}
\end{figure}

We carry out detailed investigations in the fully metallic samples. Figure~2 show the Hall resistivity data of three samples with different thicknesses across a large temperature range. All results show step-like behaviors with decreasing anomalous Hall signal at lower temperatures. Figure~3 summarizes $\rho_{AH}$ as a function of $\rho_{xx}(\mu_0 H=0)$ , showing a linear dependence for each sample. To address the relation between $\rho_{AH}$ and $\rho_{xx}$, we use a phenomenological expression $\rho_{AH}=c_0+c_1\rho_{xx}+c_2\rho_{xx}^2$to fit the data (solid curves in Fig.~3). The quadratic terms $c_2$ obtained from the fitting are 0.2 (20-nm), 4 (28-nm), -18 (45-nm) $\Omega^{-1}$~cm$^{-1}$, respectively. These values are two to three orders of magnitude smaller than those in other ferromagnetic thin films such as Fe, Co, etc.~\cite{Tian09,Hou15}, although the obtained quantities of $c_0$ and $c_1$ are comparable. The quadratic term is therefore negligible. We further obtain that $-c_0$ and $c_1\rho_0$ are almost equal (inset to Fig.~3). Essentially, the relation reads: $\rho_{AH}\propto(\rho_{xx}-\rho_0)$.

Conventionally, the AHE depends on the longitudinal resistivity following: $\rho_{AH}=b_0\rho_{xx}+b_1\rho_{xx}^2$, where $b_0$ and $b_1$ are material-dependent parameters. The first term arises from skew scattering; the second term is from side-jump scattering and the nontrivial Berry phase~\cite{Nagaosa10}. It has been demonstrated both theoretically~\cite{Crepieux01} and experimentally~\cite{Tian09} that the conventional skew scattering does not show temperature dependence. Therefore, the formula should read:  $\rho_{AH}=b_0\rho_0+b_1\rho_{xx}^2$, where only the second term $b\rho_{xx}^2$ varies with temperature. Clearly, this well-established relation cannot account for the linear dependence on $\rho_{xx}$ in our experiment.

We note that a similar behavior was reported in some other materials such as Yb$_{14}$MnSb$_{11}$ and Pt matrix embedded with Co nanoclusters~\cite{Sales08,Gerber04}. In Yb$_{14}$MnSb$_{11}$, the linear scaling appears only after subtracting a dominant quadratic term. Skew scattering with localized magnetic ions, which is different from the conventional scattering with non-magnetic impurities, was employed to explain the data~\cite{Sales08}. Such a Kondo mechanism may be  important in the Co embedded Pt as well~\cite{Gerber04}. However, the Kondo physics is clearly not applicable here, since SrCoO$_3$ is an itinerant ferromagnet.

Recently, it was proposed that the fluctuating, but locally correlated, spins contribute to the AHE~\cite{Ishizuka18}. The mechanism is unlikely to be responsible in SrCoO$_3$ either. In the proposed mechanism, the AHE is proportional to the scalar spin chirality, not to the magnetization. Moreover, the theory considers Dzyaloshinskii-Moriya (DM) interaction as the cause of scalar spin chirality, which is expected to be absent in SrCoO$_3$, since inversion centers exist at the center of the Co-Co bonds.

\begin{figure}
\centering
\includegraphics[width=86mm]{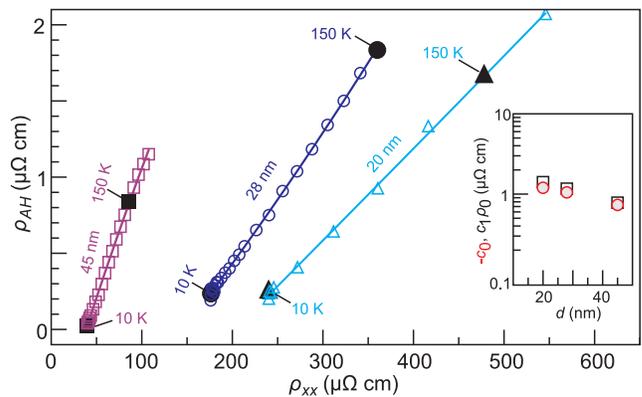}
\caption{Anomalous Hall resistivity as a function of the longitudinal resistivity at zero field. Lines are fits to the data points of each sample. Inset: fitted parameters $-c_0$ (circles) and $c_1\rho_0$ (squares) as a function of the film thickness.}
\end{figure}

\subsection{Theory\label{sec:3B}}

Theoretically, the extrinsic AHE stems from asymmetric scattering processes. The AHE at finite temperature, proportional to the magnetization, is possibly related to the vector spin chirality $\vec{S}_j\times\vec{S}_k$. When a charged non-magnetic impurity is placed into the ferromagnet, the induced electric field couples to the electric dipole of the surrounding spins. It locally breaks the inversion symmetry and causes a chiral spin fluctuation around the impurity (Fig.~4).  From the microscopic theory point of view, this is a consequence of the fact that the intermediate spin state of Co ions in SrCoO$_3$~\cite{Potze95,Zhuang98,Kunes12,Hoffmann15} allows the orbital degrees of freedom to play an important role, which may render exotic electromagnetic properties~\cite{Katsura05}.  The perturbative interaction to the spins around the impurity is:
\begin{equation}
  H_{\mathrm{imp}}\propto V_i\hat{z}\cdot\vec{S}_j\times\vec{S}_k,
\end{equation}
where $V_i$ is the impurity potential, $\vec{S}_j$ and $\vec{S}_k$ are two spins surrounding the impurity; $\hat{z}$ is the unit vector that defines the direction of the uniform magnetization. This interaction is similar to the DM interaction in noncentrosymmetric magnets except that the DM vector depends on the bond [see the Hamiltonian in Eq.~(2)]. Therefore, the impurity-induced interaction may contribute to the anomalous Hall effect by causing spin canting. To demonstrate the chiral fluctuation due to such an interaction, we consider a four-spin model that corresponds to the spins surrounding the non-magnetic impurity:
\begin{eqnarray}
  H_S&=&-J\sum_{i=1}^4 \vec{S}_{\tau(i)}\cdot\vec{S}_{\tau(i+1)}-h\sum_{i=1}^4 S_{\tau(i)}^z\nonumber\\
  &&-D\sum_{i=1}^4\left(\vec{S}_{\tau(i)}\times\vec{S}_{\tau(i+1)}\right)_z,
\end{eqnarray}
where $J$ is the Heisenberg interaction between the spins, $h$ is the magnetic field, $D$ is the impurity-induced interaction, and $\tau$: $\mathbf{Z}$$\to$ $\mathbf{Z}$ is an integer map that maps $\{1,2,3,4\}$ to the spin index of the four spins surrounding the non-magnetic impurity and $\tau(i+4)\equiv\tau(i)$; the spins are numbered by $\tau$ in the anti-clock order around the impurity. This map is introduced to avoid confusion with the later argument on MR, where we consider all the spins. Here, we ignored the contribution from other spins further away from the impurity as their canting is expected to be much smaller. The qualitative feature of our results is irrespective of the cluster shape of those spins considered. Using the classical spin-wave approximation, we find
\begin{equation}
  \langle\left(\vec{S}_{\tau(i)}\times\vec{S}_{\tau(i+1)}\right)_z\rangle=\frac{TD}{(J+h/2)^2-D^2},
\end{equation}
where $\langle\left(\cdot\cdot\cdot\right)_z\rangle$ is the thermal average of the $z$ component of the vector spin chirality. This equation indicates that the impurity-induced interactions give finite vector spin chirality only at finite temperature when the interaction is sufficiently small. Unlike the scalar spin chirality, the vector spin chirality itself does not break the time-reversal symmetry. Therefore, it is expected that the anomalous Hall conductivity is proportional to the magnetization, which is an indicator of time-reversal symmetry breaking.

\begin{figure}
\centering
\includegraphics[width=86mm]{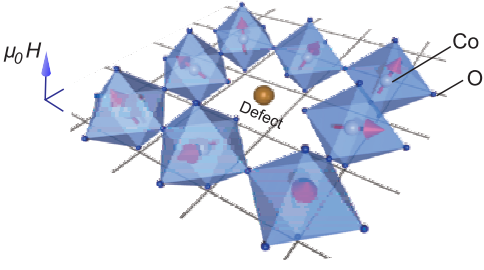}
\caption{Theoretical model of a chiral spin structure around an impurity.  Red arrows indicate the tilted spins of Co due to the presence of a central defect. Such an effect is most pronounced for the nearest neighbors. }
\end{figure}

Notably, the skew scattering often appears from the third order in the perturbation (or in the second order in Born approximation). A first Born approximation considering the scattering by two magnetic moments is insufficient. Indeed, a former study considering the vector spin chirality reported that the anomalous Hall effect related to the vector spin chirality vanishes in the bulk~\cite{Taguchi09}. Therefore, the leading order must stem from the process that involves two spins and a nonmagnetic impurity. Considering the two-spin process in Ref.~\cite{Ishizuka18} and its interference with the first-order scattering term by the non-magnetic impurity, we find the scattering amplitude from the electrons with momentum $\vec{k}$ and spin $\sigma$ to that of $\vec{k^\prime}$and $\sigma$ reads
\begin{eqnarray}
  W_{k\sigma,k^\prime\sigma}^-&=&-\sigma n_i \frac{16J_K^2V_i ma^2}{(2\pi)^7}k\langle\left(\vec{S}_{\tau(i)}\times\vec{S}_{\tau(i+1)}\right)_z\rangle\nonumber\\
  &&\cdot\left(\vec{k}\times\vec{k^\prime}\right)_z,
\end{eqnarray}
where $\sigma=\pm 1$ is the spin index of itinerant electrons, $n_i$ is the density of non-magnetic impurities, $V_i$ is the strength of the impurities, $J_K$ is the exchange coupling between the electrons and the localized moments, and $m$ is the effective mass of electrons.

In our experiment, the resistivity $\rho_{xx}$ consists of two components $\rho_{xx}=\rho_0+\rho_m$, where $\rho_0$ and $\rho_m$ are impurity and the magnetic contributions, respectively. In the Boltzmann theory, the anomalous Hall conductivity induced by the asymmetric scattering $W_{kk^\prime}^-=w(\vec{k}\times\vec{k^\prime})_z$ is: $\sigma_{xy}\propto\tau^2w\propto n_i \rho_m/\rho_{xx}^2$, where $n_i$ is the number of impurities. Here, we used the fact that $w\propto n_i \langle\left(\vec{S}_{\tau(i)}\times\vec{S}_{\tau(i+1)}\right)_z\rangle$, and $\langle\left(\vec{S}_{\tau(i)}\times\vec{S}_{\tau(i+1)}\right)_z\rangle\propto\langle(S^x)^2\rangle$ are proportional to $\rho_m$. Therefore, the Hall resistivity reads $\rho_{xy}\propto\rho_{xx}^2\sigma_{xy}\propto n_i\rho_m\sim n_i[\rho(T)-\rho(T=0)]$, qualitatively consistent with the experiment.

We further estimate the Hall angle $\theta_H\equiv\sigma_{xy}/\sigma_{xx}$ due to the vector spin chirality. We focus on the low temperature region, where the linear spin-wave approximation is accurate. We first estimate the magnitude of the impurity-induced interaction. We assume that: (1) the electric charge of the impurity is of the order of the elementary charge; (2) the scalar potential induced by the impurity has the form of the Coulomb potential; (3) the distance between the impurity and the spins are on the order of the lattice constant $a=4\times10^{-10}$~m. Taking the relative dielectric permittivity $\epsilon/\epsilon_0=10$, the model yields an electric field of $|\vec{E}|\sim 10^9$~V/m. On the other hand, the typical magnitude of the electric polarization induced by spin canting was recently studied in details for the transition-metal oxides~\cite{Jia06}; the calculation showed that the electric polarization of the form $\vec{P}=B\vec{e}_{ij}\times(\vec{S}_i\times\vec{S}_j)$ is about $B\sim 10^2$~nC/cm$^2$ for the nearest-neighbor spins. Hence, the polarization per bond reads: $\vec{p}=\vec{P}a^3\sim 10^{-31}$~C~m. By employing these results, we find the impurity-induced term to be
\begin{equation}
  H_{\mathrm{imp}}=-\vec{p}\cdot\vec{E}\sim 10^{-22}(\vec{S}_{\tau(i)}\times\vec{S}_{\tau(i+1)})J.
\end{equation}
Based on the classical spin-wave theory, we find $\langle\left(\vec{S}_{\tau(i)}\times\vec{S}_{\tau(i+1)}\right)_z\rangle\sim\frac{TD}{J^2}\sim10^{-2}$, assuming $J\sim$100~K and $T=10$~K. The magnitude of the impurity potential $V_0$ is then estimated via the first Born approximation.

From experiment, we obtain $\sigma_{xx}\sim10^6$~S/m. Using the first Born approximation, we find $\frac{1}{\tau_{\mathrm{imp}}}=\frac{n_i V_i^2}{(2\pi)^2\hbar}\rho(\varepsilon_F)$, where $\rho(\varepsilon_F)$   is the density of states (DOS) at the Fermi energy $\varepsilon_F$. From $\sigma_{xx}\sim q^2n\tau/m$, we find $\tau\sim10^{-14}$~s at $T=10$~K (Here, we ignored the contribution from the magnetic scattering, since at a sufficiently low temperature the impurity scattering dominates over magnetic scatterings.). Using electron density $n\sim 10^{29}$~m$^{-3}$, and the DOS $\rho(\varepsilon_F)\sim\frac{n}{W}\sim10^{48}$~J$^{-1}$m$^{-3}$, we find $n_iV_i^2\sim 10^{-68}$~J~m$^3$. By assuming 0.1\% density of impurity, i.e. $n_i\sim10^{24}$--10$^{25}$~m$^3$, we find $V_i\sim10^{-47}$--$10^{-46}$~J~m$^3$.

We estimate the Hall angle using the above values. In the Boltzmann theory, the Hall angle reads $\theta_H=\tau\rho(\varepsilon_F)W_{k\sigma k^\prime\sigma}^-$ where $W_{k\sigma k^\prime\sigma}^-\sim10^{-36}$~J~m$^3$/s is obtained from the second Born result assuming $k_F\sim10^{10}$~m$^{-1}$. From these results, we find $\theta_H\sim10^{-3}$--$10^{-2}$ at $T=10$~K, consistent with the experiment.

\section{Magneto-resistance}
\subsection{Experiment}
The signature of spin fluctuation can be clearly seen in the magneto-transport data. Figure~5 displays the MR of SrCoO$_3$ samples with different thicknesses at selected temperatures. These metallic samples all possess a parabolic MR (dashed curve) at low fields and a linear MR at high fields (dotted lines). The parabola show little thickness ($d$) dependence. The size effect~\cite{Suzuki98} for a negative MR can be readily excluded because otherwise the MR should depend quadratically on $d$.

Apart from the size effect, negative MR often arises due to the anisotropic magnetization of the material~\cite{Ziese02,Marrows05}. It may account for the parabolic behavior at low fields, since it becomes less distinguishable in a tilted field (see Appendix~\ref{sec:MR}). However, the contribution from the anisotropic MR (AMR) in our thin films is less than 0.5\%, which cannot account for the overall non-saturating MR seen in Fig.~5.

The domain-wall effect also produces large negative MR when sweeping from zero field. We exclude this effect since our sample shows no hysteresis and weak AMR [36], distinctly different from the expected domain-wall driven MR (see Appendix~\ref{sec:MR}). We further exclude the weak localization effect because: (1) the temperature dependent resistivity curve shows no sign of localization [Fig.~1~(c)]; (2) fitting of the magneto-conductivity with the formula for weak localization yields unphysical values (see Appendix~\ref{sec:MR}).

\begin{figure}
\centering
\includegraphics[width=86mm]{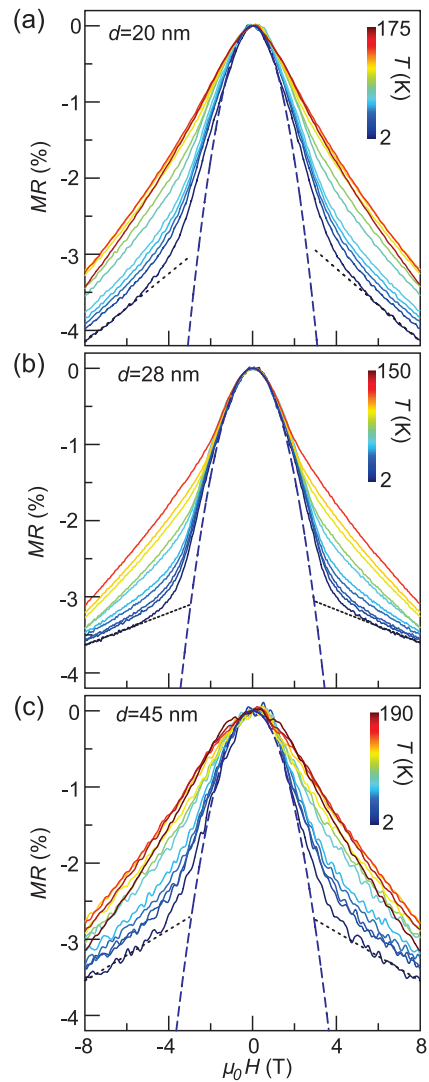}
\caption{MR of three samples with different thicknesses at selected temperatures. The dashed curves (dotted lines) are quadratic (linear) fits to the data at T=2 K at low (high) fields.}
\end{figure}

\begin{figure}
\centering
\includegraphics[width=86mm]{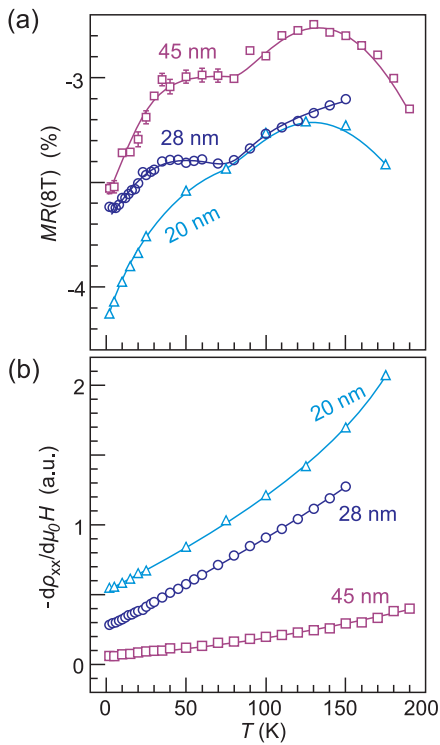}
\caption{(a) MR at 8~T as a function of temperature. MR(8T) is the mean of the expected values at $\pm$8~T, if taking linear fits to the MR in the range of $|\mu_0 H|>7$~T.  Error bar represents the standard deviation at $\pm$8~T for those linear fits. For most of the data points, the error bar is smaller than the size of the markers. (b) high field slopes of $\rho_{xx}$ for the three samples as a function of temperature.}
\end{figure}

After excluding the above-mentioned mechanisms, we attribute the observed MR to persistent spin fluctuation. First of all, the absolute value of MR becomes larger as the temperature decreases [Fig.~6~(a)]. This behavior is in sharp contrast to the conventional behavior seen in itinerant ferromagnets. There, $|\mathrm{MR}|$ is enhanced at around the Curie temperature due to spin-dependent scattering and gets suppressed at low temperatures as spins align in one direction. The unusually large $|\mathrm{MR}|$ at low temperatures in SrCoO$_3$ therefore indicates that spin-dependent scattering remains prominent. Secondly, the slope of magneto-resistivity ($d\rho_{xx}/d(\mu_0 H)$) at high fields remains finite as $T$ approaches zero [Fig.~6~(b)]. In this high field regime, the magnetization is saturated and the spin wave is expected to be significantly suppressed~\cite{Raquet02}. Previous experiments on Fe, Co and Ni thin films have demonstrated that $d\rho_{xx}/d(\mu_0 H)$ approaches zero super-linearly with decreasing temperature~\cite{Raquet02}. In contrast, our samples exhibit an almost linear decrease of $d\rho_{xx}/d(\mu_0 H)$ with a clear positive intercept as $T\rightarrow 0$.

\subsection{Theory}
To provide further insight into the effect of spin fluctuation on the resistivity, we calculate the magnetic contribution to the relaxation time using first Born approximation considering the exchange coupling $H_K=J_K\sum_i\vec{S}_i\cdot\vec{\sigma}(\vec{r}_i)$, where $\vec{\sigma}(\vec{r}_i)$ is the vector of spin operators for electron spins at $\vec{r}_i$. For the spin Hamiltonian, we consider a 3d Heisenberg model
\begin{equation}
  H_S=-J\sum_{\langle i,j\rangle}\vec{S}_i\cdot\vec{S}_j-h\sum_i S_i^z.
\end{equation}

\begin{figure}
\centering
\includegraphics[width=86mm]{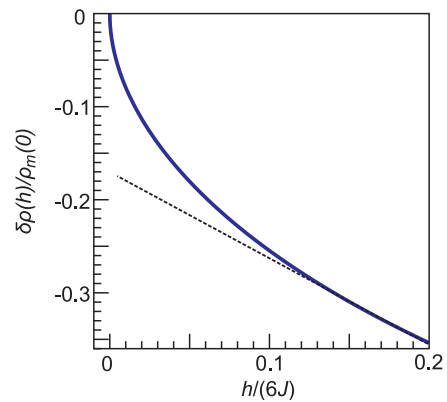}
\caption{Theoretically calculated magnetoresistance by considering the spin fluctuation. }
\end{figure}

In this section, we ignore the effective DM interaction induced by non-magnetic impurities, as they only give a higher order correction to the resistivity. We also note that, here, we consider \textbf{all} spins in the system while Sec.~\ref{sec:3B} only considers the four spins around a non-magnetic impurity. In the first Born approximation, the relaxation time $\tau_\mathrm{mag}$ reads:
\begin{equation}
  \frac{1}{\tau_\mathrm{mag}}=\frac{J_K^2\rho_\sigma(\varepsilon_{\vec{k}\sigma})}{(2\pi)^5a^3}\left[\langle(S_0^x)^2\rangle+\langle(S_0^y)^2\rangle\right],
\end{equation}
where $\varepsilon_{\vec{k}\sigma}$ is the eigen-energy for electrons with momentum $\vec{k}$ and spin $\sigma$, $\rho_\sigma(\varepsilon)$ is the density of states for electrons with spin $\sigma$ at energy $\varepsilon$, and $\langle\cdot\cdot\cdot\rangle$ represents the thermal average. The field dependence of $\tau_\mathrm{mag}$ comes from the field dependence of $\langle(S_0^x)^2\rangle$ and $\langle(S_0^y)^2\rangle$; here, we set the spin index $i=0$ assuming the translational symmetry of the ferromagnetic order. As this scattering is diagonal in the spin space, we treat the contribution from electrons with different spins independently. In deriving the above formula, we assumed that the magnetic moments are aligned along the $z$-axis, and took into account of the leading order in the fluctuation assuming the fluctuation is small. This situation applies to the high-field region where the magnetic moments are aligned almost along the field direction. In the classical spin-wave approximation, the fluctuation of spins reads
\begin{equation}
  \langle(S_0^x)^2\rangle+\langle(S_0^y)^2\rangle=\frac{T}{6J}\int_{-\pi}^\pi \frac{dk^3}{(2\pi)^3}\frac{1}{1+\eta-\Sigma_a\cos k_a},
\end{equation}
where $\eta=h/(6J)$ is the renormalized magnetic field. The sum in the integral is over the three axes $a=$x, y, z. At zero magnetic field and low temperatures (but still higher than the magnetic-field/anisotropy induced gap), the resistivity caused by spin fluctuations increases linearly with respect to $T$. Assuming $J\sim10^2$~K and $J_K\sim10^3$~K, we find $\tau_\mathrm{mag}\sim10^{-14}$~s at $T=100$~K, roughly consistent with the order of resistivity in the experiment.

Under the magnetic field, $\rho_m(h)$ is expected to be suppressed as the field pins the magnetic moments along the field direction. Within the Born approximation, the resistivity of the system follows Matthiessen's rule $\rho_{xx}=\rho_0+\rho_m$, where $\rho_i=\frac{m}{e^2 n \tau_\mathrm{imp}}$ is the contribution from the impurity scattering and  $\rho_m=\frac{m}{e^2 n \tau_\mathrm{mag}(h)}$ is the magnetic contribution; $\tau_\mathrm{imp}$ is the relaxation time for the impurity scattering.

Figure~7 plots the field dependence of the magnetoresistance $\rho_m(h)\equiv\rho_m(h)-\rho_m(0)$ renormalized by $\rho_m(h=0)$. The resistivity sharply decreases at the zero field limit, implying that the MR responds sensitively to the spin fluctuation, even when the impurity scattering is larger than the magnetic scattering. Therefore, the MR observed in the experiment is possibly related to the persistent spin fluctuation down to a very low temperature.

\section{Conclusion}
The SrCoO$_3$ thin films realized by ILG exhibit magneto-transport behaviors including: (1) the  scaling relation: $\rho_{AH}\propto(\rho_{xx}-\rho_0)$, which is distinctly different from the well-established form of $\rho_{AH}=b_0\rho_0+b_1\rho_{xx}^2$; (2) the negatively enhanced MR at low temperatures, indicating persistent spin fluctuations. We theoretically propose that impurities can induce chiral spin fluctuations in this material. By considering the local spin fluctuation around the impurity, we derive the anomalous Hall effect that is consistent with the experimentally observed relation. We further calculate the negative MR by taking into account the spin fluctuation of all spins, reproducing the non-saturating MR as seen in experiment.

\begin{figure}[!t]
\centering
\includegraphics[width=86mm]{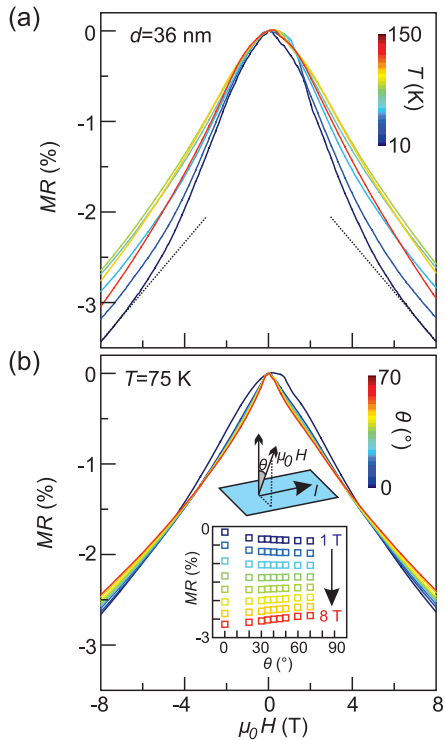}
\caption{MR and AMR of a 36-nm thick SrCoO$_3$ sample. (a) MR at a set of temperatures (10, 25, 50, 75, 100, 125, 150~K). (b) MR at 75~K in a tilted magnetic field. The magnetic field direction is perpendicular to the current (inset). Inset panel summarizes the angular dependence of MR. }
\end{figure}

\begin{figure}[!t]
\centering
\includegraphics[width=86mm]{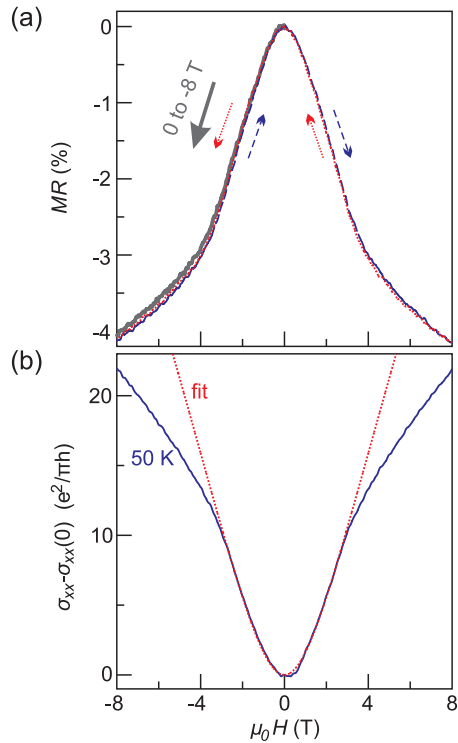}
\caption{(a) MR of the 20-nm SrCoO$_3$ thin film. The gray curve is obtained by sweeping from 0 to -8 T after the sample is zero-field cooled to 2~K. The blue and red curves are retrieved by sweeping from -8 T to 8 T and back.  (b) Magneto-conductivity at 50~K together with the fitted curve (dotted) }
\end{figure}

\acknowledgments
This study was financially supported by the National Basic Research Program of China (grants 2017YFA0304600, 2015CB921700 and 2016YFA0301004); the National Natural Science Foundation of China (grant 51561145005, 11604176); the Initiative Research Projects of Tsinghua University (grant 20141081116); and the Beijing Advanced Innovation Center for Future Chip (ICFC). H.I. and N.N. are supported by CREST JST (No. JPMJCR16F1) and JSPS KAKENHI (No. JP26103006, JP16H06717 and  JP18H04222).

Ding Zhang and Hiroaki Ishizuka contributed equally.

\appendix
\section{\label{sec:MR}Supplementary magneto-resistance data}
Figure 8 displays the magneto-transport data of a 36-nm thick thin film. Here the metallic state SrCoO$_3$ is achieved by annealing the pristine SrCoO$_{2.5}$ film in ozone~\cite{HNLee13a}. The temperature dependence of the MR is similar to that observed in Fig.~5. With this confirmation, we proceed to study the MR of this sample in a tilt magnetic field. Figure 8 (b) summarizes the data obtained at a fixed temperature but with increasing tilt angles ($\theta$). The angle $\theta$ represents the rotation of the magnetic field direction away from the normal of the sample plane. The parabolic MR at low fields disappears with increasing $\theta$. Still, the magnitude of the MR changes only slightly. As summarized in the inset, the variation of MR at each fixed field is always smaller than 0.5\%.

Figure 9~(a) shows MR of the 20-nm sample as discussed in the main text. Here we show three MR traces taken at 2~K after zero-field cooling.  These curves overlap nicely, which is in sharp contrast to the hysteretic MR caused by domain-wall effect~\cite{Ziese02,Marrows05}. Figure 9~(b) further plots the magneto-conductivity of the same sample but at 50~K. We employ the following formula to fit the data:
\begin{equation}
\sigma_{xx}-\sigma_{xx}(0)=A\frac{e^2}{\pi h}\left[\Psi\left(\frac{1}{2}+\frac{1}{x}\right)+\ln x\right],
 \end{equation}
 where $x=l_\mathrm{in}^2 \frac{4eH}{\hbar}$. This formula is adapted from the one used for two-dimensional weak localization~\cite{Scherwitzl11}. Notably, for weak localization, the prefactor $A$ is strictly 1. In contrast, we obtain $A=78$, indicating that our sample is far more conductive than that considered in the weak localization model.

\providecommand{\noopsort}[1]{}\providecommand{\singleletter}[1]{#1}%

\end{document}